\documentstyle[12pt]{article}
\newcommand{\eq}[1]{(\ref{#1})}
\newcommand{\diff}{\partial}
\newcommand{\beq}{\begin{equation}}
\newcommand{\eeq}{\end{equation}}
\newcommand{\beqn}{\begin{eqnarray}}
\newcommand{\eeqn}{\end{eqnarray}}

\def\cN{{\cal N}}
\def\cA{{\cal A}}

\def\NP{ Nucl.~Phys.}
\def\PR{ Phys.~Rev.}
\def\PL{ Phys.~Lett.}

\begin{document}

\hfill{\bf ITEP-TH-69/98} \\

\hfill{\bf hep-th/9812038}

\vspace{30mm}

\centerline{\bf \Large D-instantons probing D3-branes and}
\centerline{\bf \Large the AdS/CFT correspondence.}

\vspace{10mm}

\centerline{E.T.~Akhmedov\footnote{e--mail: akhmedov@vxitep.itep.ru}}

\centerline{Institute of Theoretical and Experimental Physics}

\centerline{Moscow, 117259, B. Cheremushkinskaya, 25.}

\vspace{10mm}

\begin{abstract}
D-instantons are considered as a probe of coinciding $N$ D3-branes. They
can feel an external metric via the commutator terms in their effective
action.  We show that when the D-instantons are separated from the D3-branes,
the metric which is probed at the one loop level, {\it exactly} coincides
with that of the BPS R-R 3-brane.  Interesting connection of this result to
the possible explanation of the AdS/CFT correspondence within IKKT M-atrix
theory is discussed.
\end{abstract}

\centerline{\bf Dedicated to the memory of Alexander Valerievich Makedonski.}

\vspace{10mm}

1. One of the ways of thinking about super-string theory is that it suggests
a method of probing the geometry of an ambient space. Until recently we
have had only one object which was able to serve for that purpose.  It was
the string itself.  At present, however, at our disposal there are D-branes
\cite{Pol95} which also can play the role of probes.

   Remarkably, the D-branes can feel a geometry at distances
smaller than the string scale \cite{KaPoDo}. This way it is possible to
probe the curved space in the vicinity of string solitons \cite{Sen,DoPoSt}.
In some circumstances that is relevant for the investigation of the SYM low
energy dynamics \cite{Wit95,Sen}.  For example, the D-branes give a geometric
description of different phenomena in the SYM theories which live on their
world-volumes (see, for example, \cite{Sen,Wit95,Geom3,Akh98}).

   Besides that, among various discoveries within the context of the D-brane
physics, there is the AdS/CFT correspondence \cite{Mal97}. It
establishes a relation between two seemingly different theories. One of them
is the four-dimensional $\cN = 4$, $SU(N)$ SYM theory, taken at large $N$ and
strong coupling ($g^2_{YM}N = g_s N >> 1$). While the other is the type IIB
SUGRA on the $AdS_5\times \left(S_5\right)_N$ background.  The latter
contains $N$ units of the elementary flux of the R-R tensor field through the
$S_5$, which is indicated by the subscript "$N$".  Also the radii of the
$AdS_5$ and $S_5$ are equivalent to each other and given by $R^4 = 4\pi
g_sN\alpha'^2$.  Here $g_s$ and $\alpha'$ are the string coupling constant
and the inverse string tension, respectively.

    Inspired by the developments of the probe physics we want to study how
it can help in the explanation of the AdS/CFT correspondence. For this
purpose we solve a warmup exercise which we hope could clear up the physics
underlying the correspondence. Following \cite{Wit97}, where a D-string was
studied as a probe of an M5-brane, we consider $n$ D-instantons probing $N$
coinciding D3-branes.  Via these tools the theory describing the M5-brane was
examined within M-atrix model \cite{Wit97}. Here we are doing the same for
the D3-brane within the IKKT picture of the M-atrix theory
\cite{Matrix,IKKT}.

   The D-instantons in the background of the D3-branes have several branches
on their "vacuum moduli space"\footnote{It is worth mentioning at this point
that in the case of D-instantons there is no such a thing as the vacuum
moduli space.  Rather the manifold in question corresponds to the
minimization of the D-instanton action {\it function}.  In fact, the
integration variables in its matrix model are just matrices rather than
fields. However, we will refer to that manifold as vacuum moduli space and to
these matrices as fields to keep an analogy with the ref.  \cite{Wit97}.}. Only
two of them are relevant for our discussion. In accordance with the ref.
\cite{Wit97}, the first one we refer to as "Coulomb branch".  It corresponds
to the case when all the D-instantons are separated from the coinciding
D3-branes. The second one is referred to as "Higgs branch" \cite{Wit97}.
It, in turn, describes the situation when all the D-instantons are sitting on
the D3-branes.  Of particular interest for our purposes are the effective
instanton actions for the "massless fields" on the branches.

    Concretely, in this note we calculate the one loop effective D-instanton
action on the "Coulomb branch". As in the ref. \cite{DoPoSt}, we want to
find the geometry which is probed in this context. Below we explain that the
D-instanton can feel the metric of a bulk space via the {\it commutator
terms} in its effective action.

    At first sight, our calculation is valid at fixed $N$ and week
coupling, then \footnote{ Hence, when $R^4/\alpha'^2 << 1$ and all the
curved part of the bulk space-time is confined inside sub-stringy distances.
Still, D-instanton can probe it.} $g^2_{YM}N = g_s N << 1$. However, because
of the aid of SUSY we exactly recover the D3-brane metric, on the branch in
question. This is the main result of our note.

   Moreover, because the dilaton of the 3-brane SUGRA solution is just a
constant, in our case it is possible to understand the singular limit of a
small distance between the probe and the D3-brane. In this limit we get the
theory describing the vicinity of the point on the D-instanton "vacuum
moduli space", which connects the "Coulomb" and "Higgs" branches. Inspired by
the ref.  \cite{Wit97}, we believe that its examination is relevant for the
explanation of the AdS/CFT correspondence. At least within IKKT M-atrix
model.

   In fact, first, the theory on the "Coulomb branch" at the singular point
can be obtained when $N\to\infty$ and $g_sN>>1$ \cite{Wit97,Mal97}.
Accordingly, {\it from our calculation} in this limit we get the D-instanton
theory on the $AdS_5\times S_5$ background.

   Second, in accordance with the ref. \cite{Wit97}, it is the latter theory
which should be equivalent to that on the "Higgs branch".  In our case this
can be traced as follows.  Really, the $n$ D-instantons on the "Higgs branch"
are equivalent to the ordinary $n$ YM instantons of the $U(N)$ gauge theory
\cite{BaGr,Green}.  At the same time it is known that the moduli space of
one $\cN = 4$ super-instanton is exactly $AdS_5\times S_5$
\cite{Witten,Kogan,Khoze,Chu}.  Moreover, as was shown in \cite{Khoze}, when
$N\to\infty$, the $n$ instanton moduli space is again the
$AdS_5\times S_5$.(All instantons tend to sit at the same point!)
Hence, the latter should be the manifold which is seen by the D-instanton
on the "Higgs branch".

   Third, it is the limit $N\to\infty$ with $n\to\infty$ taken on the "Higgs
branch", which is related to the presence of a large number of the D3-branes
in the IKKT M-atrix theory \cite{Wit97,Dou97}. Thus, even at this point we
see that the $N\to\infty$ D3-branes are equivalent to the $AdS_5\times S_5$
SUGRA.  In our next publication we hope to present a proof of those
facts and the statements of the ref.  \cite{Khoze} via the methods presented
in this note.

2.   For the beginning, we explain how a D-instanton can probe an
external metric and what is the meaning of the low energy expansion in this
context.  In early papers on the subject in question, D-branes with the world
history were considered as probes. This allowed to use the common
way of testing the metric of an external space-time \cite{DoPoSt}:  via the
kinetic part in the world-volume theory of a probe.  In the D-instanton case,
in turn, there is no such a term, because it is just a point in the ambient
space.  However, in the ref.  \cite{Dou97'} it was argued that D-branes also
can probe an external metric via commutators in their world-volume theory.
For example, in the case of $n > 1$ D-instantons one can have the following
effective action:

\beqn
S_{inst} = \frac{const \cdot \alpha'^2}{g_s} Tr \left(\phantom{\frac12} -
G_{\mu\mu'}\left(A\right) \cdot G_{\nu\nu'}\left(A\right)\cdot
\left[A^{\mu}\phantom{1^{2'}},A^{\nu}\right]
\left[A^{\mu'},A^{\nu'}\right]\right) + \nonumber \\ + \quad
{\rm super-partners} \quad + O\left(Tr\left[.,.\right]^3\right), \quad
A = Tr \sum^{10}_{\mu = 1} A^2_{\mu}. \label{inac}
\eeqn
where $A_{\mu}, \quad (\mu = 1,...,10)$ are $U(n)$ matrices. Their
eigen-values parametrize positions of the D-instantons in the target space
with the metric $G_{\mu\nu}$. The first term in this formula is unambiguous.

   The contribution $O\left(Tr\left[A,A\right]^3\right)$ in eq. \eq{inac}
includes all higher powers in commutators. They are necessary \cite{Dou97'}
to maintain an equivalent of the "general covariance"\footnote{Which is
yet to be understood \cite{Dou97'} for such a situation. However, it is
approximately respected at each order in the expansion \eq{inac}.} in this
context.  This series is an analog of an expansion in powers of derivatives,
i.e. of the low energy expansion.  In fact, under a T-duality transformation
the $A_{\mu}$ matrices are exchanged with differential operators --
the covariant derivatives $\diff_{\mu} - i \cA_{\mu}(x)$.

    Let us now describe the D-instanton theory we are going to start with.
It is the SUSY $U(n)$ matrix model with additional hyper-multiplets
describing the $N$ D3-brane background \cite{Dou97}.  Concretely, this theory
should contain hyper-multiplets both in the adjoint and $N$ in the
fundamental representations of the gauge group. Its action can be obtained,
via the reduction from four to zero dimensions, in some particular $\cN = 2$,
$U(n)$ SYM theory. In $\cN = 1$ notations the latter has the following
Lagrangian ($g_{YM}^2 = g_s$):

\beqn
L = \frac{1}{g_s} Tr Re\Bigl\{\frac12\int d^2\theta W_{\alpha}W^{\alpha} + 2
\int d^4 \theta H^+ e^{-2A} H + \nonumber \\ + 2\sum^2_{I = 1} \int d^4\theta
\Phi_I e^{-2A} \Phi_I + \frac{1}{\sqrt{2}} \int d^2 \theta H \epsilon_{IJ}
\left[\Phi_I, \Phi_J\right] + \nonumber \\ + 2 \sum^N_{p = 1} \left( \int d^4
\theta Q^+_p e^{-2A} Q_p + \int d^4 \theta \tilde{Q}_p e^{2A} \tilde{Q}^+_p
+ \frac{1}{\sqrt{2}} \int d^2 \theta \tilde{Q}_p H Q_p \right) \Bigr\},
\label{th}
\eeqn
where $A$ and $H$ are the vector-multiplet (with the super-field strength
$W_{\alpha}$) and the adjoint hyper-multiplet, respectively. They compose the
$\cN = 2$ vector-multiplet. Accordingly, $\Phi_I$ are two adjoint $\cN = 1$
hyper-multiplets, belonging to the corresponding $\cN = 2$ super-field.
Besides that, $Q_p$ and $\tilde{Q}_p$ are hyper-multiplets in the fundamental
and anti-fundamental representations of the gauge group, respectively. They,
in turn, compose the $N$ fundamental $\cN = 2$ super-fields.

   The theory \eq{th} describes the low energy dynamics on the world-volume
of the $n$ D3-branes in the unstable situation -- when $N$ D7 branes are
present. In fact, the four-dimensional theory \eq{th} has the
$\beta$-function leading to the Landau pole. Hence, it is not sensible
quantum mechanically.  However, its reduction to zero dimensions makes a
sense and describes a real and stable situation: the D-instantons in the
background of the D3-branes.

   After the reduction, the D-instanton action in question looks as follows:

\beqn
S = \frac{const \cdot \alpha'^2}{g_s} Tr \left\{\phantom{\frac12^1} -
\frac12\left[A_a,A_b\right]^2 - \left|\left[A_a,H\right]\right|^2 +
\lambda \sigma^a\left[A_a,\bar{\lambda}\right] + \right. \nonumber \\
+ \chi \sigma^a\left[A_a,\bar{\chi} \right] - i \sqrt{2}
\left(\left[\lambda,\chi\right] H^+ + \left[\bar{\lambda},\bar{\chi}\right]
H\right) + \nonumber \\
+ \frac12\left(\left[H^+, H\right] + \left[\Phi_I^+,
\Phi_I\right] + Q^+_p \otimes Q_p - \tilde{Q}_p\otimes \tilde{Q}^+_p\right)^2
- \nonumber \\
- 2 \left|\frac12\epsilon_{IJ}\left[\Phi_I,\Phi_J\right] +
\tilde{Q}_p \otimes Q_p \right|^2 -
\left|\left[A_a,\Phi_I\right]\right|^2
+ \psi_I\sigma^a \left[A_a, \bar{\psi}_I\right] - \nonumber \\ - i
\sqrt{2}\left(\left[\lambda,\psi_I\right] \Phi^+_I +
\left[\bar{\lambda},\bar{\psi}_I\right] \Phi^+_I\right)
- 2 \left|\left[\Phi_I, H\right]\right|^2 - \nonumber \\ -
\sqrt{2}\left(\left[\psi_I,\chi\right]\Phi_I +
\left[\bar{\psi}_I,\bar{\chi}\right]\Phi^+_I\right) -
\frac{1}{\sqrt{2}} \left(H \epsilon_{IJ} \left[\psi_I,\psi_J\right] +
H^+ \epsilon_{IJ} \left[\bar{\psi}_I,\bar{\psi}_J\right]\right)
\phantom{\frac12^1} - \nonumber \\ - \left|A_a
Q_p\right|^2 - \left|A_a \tilde{Q}_p\right|^2 - 2\left|H Q_p\right|^2 -
2\left|H \tilde{Q}_p\right|^2 + \nonumber \\
+ \rho_p \sigma^a A_a
\bar{\rho}_p + \tilde{\rho}_p \sigma^a A_a \bar{\tilde{\rho}}_p +
i\sqrt{2}\left(Q^+_p\lambda\rho_p - \bar{\rho}_p \bar{\lambda} Q_p -
\tilde{Q}^+_p \lambda \tilde{\rho}_p + \bar{\tilde{\rho}}_p \bar{\lambda}
Q_p\right) - \nonumber \\
- \left. \phantom{\frac12^1} \sqrt{2}\left(\tilde{\rho}_p \chi Q_p +
\tilde{Q}_p \chi \rho_p + \tilde{\rho}_p H \rho_p + \bar{\rho}_p \bar{\chi}
\tilde{Q}^+_p + Q^+_p \bar{\chi}\bar{\tilde{\rho}}_p + \bar{\rho}_p H^+
\bar{\tilde{\rho}}_p\right) \right\}. \label{ac}
\eeqn
Here $\sigma^0 = I$ and $\sigma^a, \quad (a = 1,...,3)$ are the Pauli
matrices. While $\otimes$ means the external product over the $U(n)$ color
indexes. We have suppressed them along with the spinor ones. Also from now on
we assume a summation over all repeated indexes if the inverse is not stated.

In the action \eq{ac} we have symbolized bosonic component fields in the same
way as the corresponding $\cN = 1$ super-fields.  Also $\lambda$ and $\chi$
are super-partners of the four real $A_a$ and one complex $H$ scalars.
Similarly $\psi_I$ are that of the two complex fields $\Phi_I$. While,
$\rho_p$ and $\tilde{\rho}_p$ are super-partners of the $N$ complex scalars
$Q_p$ and $\tilde{Q}_p$, respectively.  For our purposes, in deriving the
action \eq{ac}, we have assumed that the D3-branes are coinciding and placed
at the origin of the six-dimensional transverse space.

Let us explain now the string and geometric meanings of those bosonic fields.
The ten hyper-multiplets $A_a, \quad (a = 1,...,4)$, $H = \frac{1}{\sqrt{2}}
\left(A_5 + i A_6\right)$ and $\Phi_I, \quad (I = 1,2)$ are coming from the
strings attached by both their ends to the D-instantons. According to the
ref. \cite{Itoyama} one can relate the theory \eq{ac}, taken
only for these fields, to the D-instanton action in the flat target space.
At the same time the $N$ hyper-multiplets $Q_p$ and $\tilde{Q}_p$ are coming
from the strings stretched between the D-instantons and the D3-branes.

   Thus, the eigen-values of the six scalars $A_m, \quad (m = 1,...,6)$
parametrize relative positions of the $n$ D-instantons to the D3-branes.
When those fields acquire non-zero "VEVs", the former are separated from the
latter.  At the same time, the eigen-values of the four scalars $\Phi_I$
represent positions of the D-instantons in the four directions tangential to
the D3-branes.

  To understand the geometric meaning of the $Q_p$ and $\tilde{Q}_p$ fields
one can look for the SUSY "vacuum state" of the theory \eq{ac}. Among the
conditions on the state there are as follows:

\beqn
\left[\Phi_I^+,\Phi_I\right] + Q^+_p \otimes Q_p - \tilde{Q}_p\otimes
\tilde{Q}^+_p = 0 \nonumber \\
\frac12\epsilon_{IJ}\left[\Phi_I,\Phi_J\right] + \tilde{Q}_p \otimes
Q_p = 0.
\eeqn
They, along with the factorization over the $U(n)$ gauge
group, represent the Hyper-Kahler reduction procedure. Which is the main
ingredient in the ADHM construction of the $n$ instanton moduli space in the
$U(N)$ gauge theory.  (See, for example, \cite{Douglas,Wit97} for the
discussion on this point in the context related to our case.) Therefore,
while eigen-values of $\Phi_I$ correspond to the positions of the $n$
instantons, their sizes and relative color orientations are represented by
the fields $Q_p$ and $\tilde{Q}_p$.  If the latter acquire non-zero "VEVs",
the D-instantons are confined to the D3-brane world-volume. They become
ordinary YM instantons \cite{BaGr} with the sizes set by the background of
$Q_p$ and $\tilde{Q}_p$.

3. Now we will be looking for the D-instanton effective action on the
"Coulomb branch". As was explained above, it is enough for us to consider
the terms which are of the quadratic order in commutators.

On the branch of interest we have a non-zero "VEV" for the $\sum^6_{m = 1}
A^2_m \equiv |A|^2$.  Via the gauge fixing it can be set to be (no summation
over $i$ is assumed in this formula):

\beq
\left(\sum^6_{m = 1} A^2_m\right)_{ij} \equiv \left|A\right|^2_{ij} =
\frac{r^2_i}{\alpha'^2} \delta_{ij}, \label{VEV}
\eeq
where $r_i, \quad (i = 1,...,n)$ are the distances between the D-instantons
and the D3-brane.  Eventually, to simplify an interpretation of the final
result, we will take all $r_i \to r$ (see below).  At the same time, while
$|A|^2$ is diagonal, $A_m$ themselves are assumed to have small non-diagonal
parts.  Also we consider non-zero background $\Phi_I$ fields with the similar
conditions on their non-diagonal parts. Thus, we slightly break SUSY and have
small but non-zero commutators of the matrices $A_m$ and $\Phi_I$, which is
similar to the case of a non-zero velocity of a probe brane as in the ref.
\cite{DoPoSt}.

  As we see from the eq. \eq{ac}  the fields $Q_p$, $\tilde{Q}_p$, $\rho_p$
and $\tilde{\rho}_p$ acquire "masses" (set by $r_i$) in the background
\eq{VEV}.  We integrate them out to get the effective action for the
"massless" fields $A_m$ and $\Phi_I$.  While "massless" fermions $\lambda$,
$\chi$ and $\psi_I$ are set to zero.

   For the background in question the action \eq{ac} can be rewritten as
follows:

\beqn
S = S_{cl} + \frac{const \cdot \alpha'^2}{g_s} Tr \left\{
\phantom{\frac12} F F^+ -
\sqrt{2} F \tilde{Q}_p \otimes Q_p  - \sqrt{2} F^+ \tilde{Q}^+_p \otimes Q^+_p
- \right. \nonumber \\ - \frac12 D^2 + D \left(Q^+_p\otimes Q_p -
\tilde{Q}_p\otimes \tilde{Q}^+_p\right) - \nonumber \\ - \left[\Phi^+_I,
\Phi_I\right] \left(Q^+_p\otimes Q_p - \tilde{Q}_p\otimes
\tilde{Q}^+_p\right) - \epsilon_{IJ} \left[\Phi_I, \Phi_J\right]
\tilde{Q}^+_p \otimes Q^+_p - \nonumber \\ - \epsilon_{IJ} \left[\Phi^+_I,
\Phi^+_J\right] \tilde{Q}_p \otimes Q_p - \left|A_m Q_p\right|^2 - \left|A_m
\tilde{Q}_p\right|^2 + \rho_p \sigma^a A_a \bar{\rho}_p + \tilde{\rho}_p
\sigma^a A_a \bar{\tilde{\rho}}_p - \nonumber \\ \left. \phantom{\frac12} -
\sqrt{2}\left(\tilde{\rho}_p H \rho_p + \bar{\tilde{\rho}}_p H^+
\bar{\rho}_p\right)\right\},
\eeqn
where $S_{cl}$ is that part of \eq{ac} which depends {\it only} on the $A_m,
\quad (m = 1,...,6)$ and $\Phi_I$ fields and we have explicitely wrote all
the terms depending on the fundamental hyper-multiplets.  The fields $D$, $F$
and $F^+$ in this formula can be integrated out, leading to the {\it quartic}
in $Q_p$ and $\tilde{Q}_p$ contributions from the eq. \eq{ac}.

   After a straightforward Gaussian integration over the fundamental
hyper-multiplets and simple manipulations with the logarithms we get the
following effective action\footnote{We do not include the Fadeev-Popov
determinant into this expression, because after the gauge fixing
\eq{VEV} it gives no contribution to the terms quadratic in commutators.}:

\beqn
S_{eff} \approx \frac{const \cdot \alpha'^2}{g_s} Tr \left\{-
\frac12\left[A_m,A_n\right]^2 - \left|\left[A_m,\Phi_I\right]\right|^2 +
\frac12 \left[\Phi^+_I,\Phi_I\right]^2 - \right. \nonumber \\ \left. -
\frac12 \left|\epsilon_{IJ}\left[\Phi_I,\Phi_J\right]\right|^2 + F F^+ -
\frac12 D^2 \right\} - \nonumber \\ - \frac{N}{2} Tr \log
\left\{1 + \frac{\alpha'^4}{r^4}\left(\left[\Phi^+_I,
\Phi_I\right] - D\right)^2 - \frac{\alpha'^4}{r^4}
\left|\epsilon_{IJ}\left[\Phi_I, \Phi_J\right] + \sqrt{2}F\right|^2 +
...\right\} + \nonumber \\ + \frac{N}{2} Tr \log \left\{1 -
\frac{\alpha'^4}{r^4}\left[A_m, A_n\right]^2 + ...\right\} + ..., \label{det}
\eeqn
where dots stand for higher powers in the commutators and fields $F$ and $D$.
We have explicitely substituted into this formula the $S_{cl}$, "VEV" of the
$|A|^2$ from the eq.  \eq{VEV} and have taken the limit $r_i \to r$. The
first logarithm, in the obtained expression, is the contribution of the
bosons $Q_p$ and $\tilde{Q}_p$, while the second one is due to the fermions
$\rho_p$ and $\tilde{\rho}_p$.

   To get eventually the terms which are quadratic in commutators,
one can expand the logarithms from the eq. \eq{det} only to the leading order.
In fact, all higher corrections give, after the integration over the $F$ and
$D$ fields, contributions to higher terms in the series \eq{inac}.  Thus,
performing the expansion and integration over the $F$ and $D$ fields, we
get the ultimate answer:

\beqn
S_{eff} \approx \frac{const \cdot \alpha'^2}{g_s} \left\{- \frac12 \left(1
+ \frac{R^4}{r^4}\right) Tr \left[A_m,A_n\right]^2 -
Tr\left|\left[A_m,\Phi_I\right]\right|^2 +
\phantom{\left(\frac{1^4}{2^4}\right)^{-1}} \right. \nonumber
\\ \left. + \frac12 \left(1 + \frac{R^4}{r^4}\right)^{-1}
Tr\left(\left[\Phi^+_I,\Phi_I\right]^2 -
\left|\epsilon_{IJ}\left[\Phi_I,\Phi_J\right]\right|^2\right) \right\} +
O\left(Tr\left[.,.\right]^3\right), \label{ult}
\eeqn
which is exactly the D-instanton effective action \eq{inac} in the
BPS 3-brane background:

\beq
d^2s = \left(1 + \frac{R^4}{r^4}\right)^{-\frac12} \sum^2_{I=1}d\varphi_I
d\bar{\varphi}_I + \left(1 + \frac{R^4}{r^4}\right)^{\frac12} \sum^6_{m=1}d
a_m d a_m, \label{metr}
\eeq
where $r^2 = \sum^6_{m=1} a^2_m$. Now one can see that when all commutators
are equal to zero, i.e. we respect the SUSY, the bosonic and fermionic
determinants are perfectly cancel between each other. Moreover, because of
the general SUSY arguments, the eq.  \eq{ult} is an exact result at this order
of the expansion \eq{inac}.

4. As we see the D-instantons allow to probe directly an external metric
as in the eq. \eq{inac} rather than its implicit form, coming from the
Born-Infeld action as in \cite{DoPoSt}. In the latter case one is able to
fix an analog of the "Coulomb behavior" ($\sim v^2/r^2$) in the expansion of
an effective action in powers of the velocity ($v$).  Which is more or less
expectable within the SYM theory.  But this raises the question:  how one can
understand, within this framework, such a global characteristic of the bulk
space-time as the horizon? We could hope that simplicity of the D-instanton
matrix model, combined with the direct probing of the metric \eq{inac}, can
help in obtaining the horizon directly in the effective action.

5. Author is indebted for valuable discussions to A.~Gorsky, A.~Losev,
A.~Morozov, I.~Polyubin and especially to A.~Gerasimov. This work was
partially supported by RFBR 96-021-7230, RFBR 98-02-16575, INTAS 96-538 and
RFBR-96-15-96740 for scientific schools.


\begin{thebibliography}{50}

\bibitem{Pol95} J.~Polchinski, {\it ``TASI Lectures on D-branes''},
hep-th/9611050; \\
J.~Polchinski, S.~Chaudhuri and C.~Johnson, {\it ``Notes on
D-branes''}, hep-th/9602052.

\bibitem{KaPoDo} M.~Douglas, D.~Kabat, P.~Pouliot and S.~Shenker, {\it \NP},
{\bf B485} (1997) 85.

\bibitem{Sen} A.~Sen, {\it \NP}, {\bf B475} (1996) 562; {\it \NP}, {\bf B489}
(1997) 139; {\it \NP}, {\bf B498} (1997) 135;\\
T.~Banks, M.~Douglas and N.~Seiberg, {\it \PL}, {\bf B387} (1996) 278;\\
M.~Douglas, D.~Lowe and J.~Schwarz, {\it \PL}, {\bf B394} (1997) 297.

\bibitem{DoPoSt} M.~Douglas, J.~Polchinski and A.~Strominger, {\it ``Probing
Five-Dimensional Black Holes with D-branes''}, hep-th/9703031.

\bibitem{Wit95} E.~Witten, {\it \NP}, {\bf B460} (1996) 335.

\bibitem{Geom3} M.R.~Douglas and G.~Moore, {\it ``D-branes, Quivers and ALE
Instantons''}, hep-th/9603167.

\bibitem{Akh98} E.~Akhmedov, {\it "A remark on the AdS/CFT correspondence and
the renormalization group flow"}, hep-th/9806217, accepted for publication in
PLB.

\bibitem{Mal97} J.M.~Maldacena, {\it ``The Large N Limit of Superconformal
Field Theories and Supergravity''}, hep-th/9711200.

\bibitem{Wit97} E.~Witten, {\it J.High Energy Phys.}, {\bf 07} (1997) 003.

\bibitem{Matrix} T.~Banks, W.~Fischler, S.~Shenker and L.~Susskind,
{\it \PR}, {\bf D55} (1997) 5112.

\bibitem{IKKT} N.~Ishibashi, H.~Kawai, Y.~Kitazawa and A.~Tsuchiba, {\it
\NP}, {\bf 498} (1997) 467.

\bibitem{Dou97'} M.~Douglas, {\it \NP Proc.Suppl.}, {\bf 68} (1998) 381.

\bibitem{BaGr} T.~Banks and M.~Green, {\it J.High Energy Phys.}, {\bf 05}
(1998) 002.

\bibitem{Green} M.~Bianchi, M.~Green, S.~Kovacs and G.~Rossi, {\it J. High
Energy Phys.}, {\bf 9808} (1998) 013.

\bibitem{Witten} E.~Witten, {\it J. High Energy Phys.}, {\bf 9807} (1998)
006.

\bibitem{Kogan} I.~Kogan and G.~Luzon, {\it "D-instantons on the boundary"},
hep-th/9806197.

\bibitem{Khoze} M.~Dorey, T.~Hollowood, V.~Khoze, M.~Mattis and S.~Vandoren,
{\it "Multi-Instantons and Maldacena's Conjecture"}, hep-th/9810243.

\bibitem{Chu} C.-S.~Chu, P.-M.~Ho and Y.-Y.~Wu, {\it "D-Instanton in $AdS_5$
and Instanton in $SYM_4$"}, hep-th/9806103.

\bibitem{Dou97} M.~Douglas and M.~Berkooz, {\it \PL}, {\bf B395} (1997) 196.

\bibitem{Itoyama} H.~Itoyama and A.~Tokura, {\it \PR}, {\bf D58} (1998)
026002.

\bibitem{Douglas} M.Douglas, {\it "Branes within Branes"}, hep-th/9512077.

\end{thebibliography}
\end{document}